\documentclass[prl,twocolumn]{revtex4}

\begin{document}
\title{Correlation additivity relation is superadditive for separable state}

\author{Zhan-jun Zhang}
\affiliation{School of Physics \& Material Science, Anhui
University, Hefei 230039, China}

\maketitle
How to characterize and quantify correlations in a quantum system 
has attracted much attention so far. Recently K. Modi et al\cite{1} 
put forward a unified view on quantum and classical correlations.
This is a very important contribution to correlation study for it 
allows to put all correlations on an equal footing and offers a quite 
concise and clear physics picture (see fig.1 in Ref.[1]
for illustration). Specifically, they defined several derivative states 
of a given state and used their relative entropies to define various 
correlations. To well demonstrate their unified view, K. Modi et al[1] 
showed three examples, where all correlations were 
calculated and their additivity relations were found. The first example
is about a simple family of two-qubit mixed states, i.e.,
Bell-diagonal states, which comprise both separable and 
entangled states. The second and third examples aim at the three-qubit 
W state and the four-qubit cluster state, which represent multipartite 
entangled pure states. Following the three examples, K. Modi et al[1]
further wrote: {\it "From the examples above we {\bf conjecture} that the correlations 
of a quantum state are subadditive in the sense $T_\rho \ge E+Q+C_\sigma$. 
The source of the subadditivity may be due to entanglement being less than 
the difference in the entropies of $\rho$ and $\sigma$, i.e., 
$S(\sigma) \ge -{\rm tr}\rho \log \sigma.$  We have not been able to prove 
this explicitly nor have we found an example showing the contrary."} \ \  
{\bf Is the conjecture right? } Specifically, whether the correlation subadditivity 
relation $T_\rho \ge E+Q+C_\sigma$ holds for any state? In this paper we will 
show an opposite example to deny the generality of the conjecture
for all states. We will verify the correlation additivity relation is superadditive instead of subadditive
for separable states. We think, this work is helpful for those who are
still working hard to explicitly prove the subadditivity relation and to find its source.

For simplicity of description we will use the same definitions and symbols 
as those in Ref.[1] hereafter. For all the states in Modi et al's three examples the 
subadditivity relation $T_\rho \ge E+Q+C_\sigma$ holds indeed, particularly the first 
example including some separable states. Based on them, Modi et al further 
conjectured that the subadditivity relation holds for any quantum state.
Is it right? Let us generally consider $\rho$ is separable. In this case,  
its closest separable state $\sigma$ is naturally itself, i.e., $\sigma=\rho$,
then the quantum entanglement $E$ of $\rho$ is zero and its quantum mutual 
information $T_{\rho}$ becomes $T_{\sigma}$. As a consequence, the conjectured subadditivity 
relation is converted into $T_{\sigma} \ge Q+C_\sigma$. In fact, Modi et al[1] 
have already derived an important equality $T_{\sigma} = Q+C_\sigma-L_\sigma$. 
Hence, one is readily to conclude that, if $T_{\sigma} \ge Q+C_\sigma$ holds 
for any separable state $\sigma$, then $L_\sigma \equiv 0$ and $T_{\sigma} \equiv Q+C_\sigma$, 
because all the quantities here including $L_\sigma$ are relative entropies 
and should be nonnegative. Incidentally, in the first example\cite{1} 
some separable states are definitively included and their $L$'s are strictly 
zero indeed. Nonetheless, whether $L$ is constantly zero for any separable state? 
If yes, then the part about separable states in the figure 1 of Ref.[1] should be 
revised at least. Unfortunately, the answer is negative. Consider the separable state $\rho=\sigma=\frac{1}{2}(|00\rangle\langle 00| + |1H\rangle\langle1H|)$, 
where $|H\rangle=(|0\rangle+|1\rangle)/\sqrt2$. By strict deductions, one can get 
its closest classical state $\chi_{\sigma}=\frac{1}{2}|00\rangle\langle 00|+\frac{1}{4}|10\rangle\langle 10|+\frac{1}{4}|11\rangle\langle 11|$. Note that $\chi_{\sigma}$
is determined completely in virtue of the minimization defined by the equation 3 of Ref.[1], 
that is, minimizing the quantum dissonance $Q$. Therefore, $Q$ is simultaneously 
determined after the minimization, i.e., $Q= \frac{1}{2}=0.5$. Easily one can get 
the product states $\pi_{\sigma}= \frac{1}{2}(|0\rangle\langle 0|+|1\rangle\langle 1|)\otimes (\frac{3}{4}|0\rangle\langle 0|+\frac{1}{4}|1\rangle\langle 1| +\frac{1}{4}|1\rangle\langle 0| +\frac{1}{4}|0\rangle\langle 1|)$ and $\pi_{\chi_{\sigma}}=\frac{1}{2}(|0\rangle\langle 0|+|1\rangle\langle 1|)\otimes (\frac{3}{4}|0\rangle\langle 0|+\frac{1}{4}|1\rangle\langle 1|)$.
In terms of Modi et al's definitions, further one can get the quantum mutual information
$T_\sigma=1-\frac{1}{4} [(2+\sqrt 2)\log_2 (2+\sqrt 2) + (2-\sqrt 2) \log_2 (2-\sqrt 2)]\approx 0.601$,
the classical correlation $C_{\sigma}= \frac{3}{4} \log \frac{4}{3} \approx 0.311$,
and the quantity $L_{\sigma}=1-\frac{3}{4}\log 3+ \frac{1}{4} [(2+\sqrt 2)\log_2 (2+\sqrt 2) + (2-\sqrt 2) \log_2 (2-\sqrt 2)]\approx 0.210$. Obviously, $L_{\sigma} >0$ and $T_{\sigma} < Q+C_\sigma$. This indicates that the
conjectured subadditivity relation $T_\rho \ge E+Q+C_\sigma$ is not a general conclusion and does not always
hold for all quantum states. Since $L_{\sigma} \geq 0$, 
the correlation additivity relation for all separable states is actually superadditive instead, i.e., $T_{\sigma} \le Q+C_\sigma$. By far, it is easy to see that,
to strictly prove the correlation subadditivity relation for any state 
and to find its source are fruitless.

ZJ is very grateful to Prof. V. Vedral for his discussion during ZJ's visit 
and his admission on this work via our late private communication after reading 
the detailed deductions. This work is supported by the Specialized Research Fund for the Doctoral 
Program of Higher Education under Grant No. 20103401110007, the National Natural 
Science Foundation of China under Grant No. 10975001, and the 211 Project of Anhui University.


\begin{thebibliography}{99}

\bibitem{1}  K. Modi, T. Paterek, W. Son, V. Vedral, M. Williamson, Phys. Rev. Lett. {\bf 104}, 080501 (2010).
\end{thebibliography}
\end{document}